\documentclass{aa}
\usepackage{graphicx}
\usepackage{astrobib}
\usepackage{journals}

\begin{document}

\title{Population X: Are the super-Eddington X-ray sources beamed jets
in microblazars or intermediate mass black holes?}

%\subtitle{}

\author{Elmar K\"ording, Heino Falcke \and Sera Markoff\thanks{Humboldt research fellow}}

\institute{ Max-Planck-Institut f\"ur  Radioastronomie,
  Auf dem H\"ugel 69, D-53121 Bonn, Germany }

\date{Astronomy \& Astrophysics, in press (2002)}

\titlerunning{Population X}
\authorrunning{K\"ording et al.}

\abstract{Recent X-ray observations reveal an increasing number of
X-ray sources in nearby galaxies exceeding luminosities of $L_{\rm
x}\ga2 \cdot 10^{39} \mbox{erg}\ \mbox{s}^{-1}$.  Assuming isotropic
emission, the Eddington limit suggests a population of
intermediate-mass black holes of $M_\bullet\gg10M_\odot$. However,
\citeN{MarkoffFalckeFender2001} proposed that jets may be
contributing to the X-ray emission from X-ray binaries (XRBs), implying that
some X-ray sources may be relativistically beamed.  This could reduce
the required black hole masses to standard values. To test this
hypothesis, we investigate a simple X-ray population synthesis model
for X-ray point sources in galaxies with relativistic beaming and
compare it with an isotropic emission model.  The model is used to
explain a combined data set of X-ray point sources in nearby
galaxies. We show that the current distributions are consistent with
black hole masses $M_\odot\la10$ and bulk Lorentz factors for jets in
microquasars of $\gamma_{\rm j}\sim5$.  Alternatively, intermediate
mass black holes up to 1000 $M_\odot$ are required which are 
distributed in a powerlaw with roughly $\frac{dN}{dM} \sim M^{-2}$.  
}

\maketitle

\section{Introduction}
In recent years X-ray observations have revealed several off-nucleus
ultra-luminous X-ray sources (ULXs) in the luminosity range $10^{39}\
- \ 10^{40}\mbox{erg}\ \mbox{s}^{-1}$ within nearby galaxies 
(\citeNP{LaParolaPeresFabbiano2001}; \citeNP{MizunoKubotaMakishim2001}; 
\citeNP{BauerBrandtSambruna2001}; \citeNP{ColbertMushotzk1999}).  
The Eddington limit for
an accreting object with mass $M$ is $L_{\rm Edd} \approx 1.25 \cdot
10^{38} \frac{M}{M_\odot} \mbox{erg}\ \mbox{s}^{-1}$, which implies
that these sources are super-Eddington for stellar mass objects.  Some
ULXs show spectral transitions from a soft spectrum to a hard power
law and luminosity variability (e.g. \citeNP{MizunoKubotaMakishim2001}; 
\citeNP{KubotaMizunoMakishima2001}), ruling out supernova
remnants and supporting the idea that ULXs can be attributed to
accreting black holes. To achieve the observed X-ray luminosities with
isotropically radiating accretion disks should require a population of
intermediate-mass black holes of $50-500 M_\odot$.  As discussed in
\citeN{KubotaMizunoMakishima2001}, however, the measured inner-disk
temperatures of ULXs ($T_{\rm{in}} = 1.0-1.8$ keV) are too high for
these masses and there is no established formation scenario for such
intermediate-mass black holes.

The problems with isotropic emission models have already been
discussed by \citeN{KingDaviesWard2001},
where the authors propose some form of anisotropic emission as an
alternative. A beaming factor of ten already reduces the required mass
of the black holes to expected values, but this is difficult to
achieve with pure disk models.  Recently \citeN{MarkoffFalckeFender2001} 
suggested that the spectrum of some XRBs
could be explained by a coupled disk/jet model, where some of the
X-ray emission is produced by synchrotron and inverse-Compton
radiation in a jet. This emission would naturally be relativistically
beamed. \citeN{MirabelRodriguez1999} 
(see also \citeNP{ReynoldsLoanFabian1997}) 
have pointed out that a number of nearby galaxies should host
microblazars - microquasars with relativistically beamed jets pointed
towards the observer. We will here investigate whether such
populations of microblazars or intermediate mass black holes can
indeed explain current data on ULXs and constrain the basic parameters
required for these models.

\section{Simple models}
\subsection{The Jet/Disk Model}
Black hole candidate XRBs can exist in several states, the two most
distinct of which are a high/soft state where the observed spectrum is
soft and thermally-dominated, and a low/hard state dominated by a
non-thermal hard power law spectrum (e.g. \citeNP{Nowak1995}). Which
state an XRB is in seems to depend on the accretion rate. One scenario
for the evolution of XRBs is that the inner part of the accretion disk
consists of an optically thin, advection-dominated accretion flow
(ADAF) existing up to a transition radius where the accretion flow
turns into a standard \cite{ShakuraSunyaev1973} optically thick
disk \cite{EsinMcClintockNarayan1997}.  The low/hard state seems to be
accompanied by persistent radio jets with optically thick synchrotron
emission extending up to the optical and near-infrared \cite{Fender2001}. 
Synchrotron and inverse Compton emission from the jet
could also be produced in the X-rays for low/hard and high/soft states
(\citeNP{MarkoffFalckeFender2001}; \citeNP{MarkoffFalckeFender2001b}).

For a self-consistent population synthesis model, we will have to take
 the jet and the disk separately into account. For simplicity we make
 some assumptions in order to calculate the relative importance of the
 two processes for the overall distribution of XRBs. These are:

\begin{itemize}
\item We only consider two main populations: neutron stars of mass $1.4 M_\odot$ and black
holes within a mass range of 5-15 $M_\odot$, where we
fix the amount of active black holes ( $ > 5 \cdot 10^{36} \mbox{erg} \ \mbox{s}^{-1}$) to 13\% of the number of active 
neutron stars (e.g. \citeNP{TanakaLewin1995}).
\item While the spectral states are defined only for black holes, for
simplicity we apply them also to neutron stars, and consider only the
low/hard and soft/high states.
\item The probability that a given XRB has the accretion rate 
$\dot{M}$ is given by $\mathcal{W}(\dot{M}) \sim \dot{M}^\xi $
which we assume as a power law with a
cutoff representing the Eddington limit.
\item The distribution of accretion rates and basic jet parameters
are assumed to be identical for neutrons stars and black holes.
\item Soft X-ray emission is produced by an isotropically radiating
disk and a relativistically beamed jet as discussed below.
\end{itemize}

The crucial point for a simple population synthesis model including
jet emission is how a specific accretion rate translates into an X-ray
luminosity. Here we assume that the transition between the low and the
high state happens at a critical accretion rate $\dot{M}_C$ and the
luminosities scale as follows: above $\dot{M}_C$ the disk luminosity
increases linearly with $\dot{M}$ as expected for a standard accretion
disk. Below $\dot{M}_C$, the disk luminosity increases with
$\dot{M}^2$ as expected for optically thin ADAFs (\citeNP{NarayanYi1995}; 
for a constant $\alpha$-parameter). Assuming that the
jet power scales linearly with $\dot{M}$, the optically thick jet
synchrotron emission will scale roughly as 
$L_{\rm x,jet}\propto\dot{M}^{1.4}$ 
(\citeNP{FalckeBiermann1995,FalckeBiermann1999}). 
For simplicity we assume that this scaling also
holds for the optically thin flux. At high accretion rates this
scaling must break down when a significant fraction of the jet power
is radiated away.  In this phase the radiated power can only increase
linearly with jet power. In the jet model of 
\citeN{MarkoffFalckeFender2001} and \citeN{MarkoffFalckeFender2001b} 
this happens in the high
state, roughly at $\dot M>\dot M_{\rm C}$, where the jet is
inverse-Compton cooled (radiating soft X-rays) by the accretion disk.
However, since models for the contribution of jets to the high state of
XRBs are not yet very well developed, we simply fix the luminosity of
the jet at $L_{\rm{jet}} = \eta L_{\rm{disk}}$ at $\dot{M}_C$,
where $\eta$ is a free parameter.

Hence, we use the following simple parameterization for the soft X-ray
luminosity of accretion disk and jet:

% (see also Fig.~\ref{LxMd}):

\begin{eqnarray}
L_{\rm{disk}} & =  & \left\{ 
\begin{array}{c}
\epsilon \left(\frac{\dot{M}}{\dot{M}_C}\right) \dot{M} c^2 \ \ \mbox{if} \ \ \dot{M} < \dot{M}_C  \\
\epsilon \dot{M} c^2  \ \ \mbox{if}  \ \ \dot{M}_C < \dot{M} <\dot{M}_{\rm{Edd}} 
\end{array}
\right. \qquad \nonumber \\
L_{\rm{jet}} & = &  \left\{ 
\begin{array}{c}
\eta \epsilon \left(\frac{\dot{M}}{\dot{M}_C}\right)^{0.4} \dot{M} c^2 \ \ \mbox{if} \ \ \dot{M} < \dot{M}_C  \\
\eta \epsilon \dot{M} c^2  \ \ \mbox{if} \ \ \dot{M}_C< \dot{M} <\dot{M}_{\rm{Edd}} 
\end{array}
\right.  \label{emitlaw}
\end{eqnarray}
In the following we set the radiative efficiency of the standard
accretion disk to the canonical value of $\epsilon = 0.1$.  For a
given mass $M$ the parameter $\dot{M}_{\rm{Edd}}$ has been chosen such
that the luminosity of the disk and the jet integrated over all angles
is equal to $L_{\rm Edd}$.  For simplicity, the mass distribution of
black holes is given by $d N/dM=\mathcal{V} (M)=$const.

%\begin{figure}
%\resizebox{\hsize}{!}{\includegraphics{plot1.eps}}
%\caption{Schematic logarithmic plot of the dependence of the X-ray
%luminosity as a function of the accretion rate}
%\label{LxMd}
%\end{figure}

For a bulk Lorentz factor of the jet of $\gamma_{\rm j}>1$ the jet
emission depends on the angle to the line of sight as given by 
\citeN{LindBlandford1985}. If the emission in the rest frame of the jet
follows a power-law with spectral index $\alpha$, the observed
emission is proportional to $\delta^{2+\alpha}$, where the Doppler
factor $\delta = \frac{1}{\gamma_{\rm j} (1-\beta \cos \Theta )}$.
The probability of seeing an object with an emission exceeding $L$
when in the rest frame the jet emits $L_{\rm{loc}}$ is:
\[
P(L,L_{\rm{loc}}) = \frac{1-\beta}{\beta}\left( 
\left( \frac{L_{\rm{max}}}{L} \right)^{\frac{1}{2+\alpha}} -1 \right)
\]
where $L_{\rm{max}} = \delta^{2+\alpha}(\Theta = 0) L_{\rm{loc}}$ is
the maximum possible emission.  To derive this we only consider the
jet component pointing towards us and then integrate over all 
inclination angles. Since we only discuss jets with
$\gamma_{\rm j} >2$, the emission of the counter-jet is largely
negligible.

With this parameterization, the contribution of a single population of
XRBs at a given accretion rate and mass has three parameters
$(\gamma_{\rm j}, \dot{M}_C, \eta)$, which are reasonably well
constrained by the underlying models. Observations of microquasars
show that the typical Lorentz factors are in the range $\gamma_{\rm j}
\simeq2-5$ (\citeNP{MirabelRodriguez1999}; \citeNP{FenderGarringtonMcKay1999}). The typical values for the critical accretion
rate discussed in the literature are around $\dot{M}_C\sim0.1$
\cite{NarayanYi1995} and we keep this parameter fixed. The
jet efficiency can in principle be fairly high, but
probably $ \eta \la0.3$ (\citeNP{FalckeBiermann1995} \& 1999).

At high accretion rates, the jets emits a factor $\eta$ less radiation
in its rest frame compared to the disk, but beaming will lead to an
amplification of the jet with respect to the disk for small
inclination angles.  Beaming is strongest for sources whose jets point
within the beaming cone with half-opening angle of $1/\gamma_{\rm j}$
towards the observer. For example, already $\gamma_{\rm j} = 3$ will
beam the jet component in a fraction of $5.5\%$ of the binaries by a
factor of 20. This is more than enough to make up for the less
efficient emission mechanism.  For $\gamma_{\rm j} = 5$ a
fraction of $2\%$ binaries are beamed by a factor of 77.  Therefore,
in the low-luminosity regime ($<< \epsilon \eta \dot{M}_C$) jets
should dominate (because of low radiative efficiency of ADAFs), and
again dominate in the super-Eddington regime due to  beaming.  In
the intermediate regime up to $L_{\rm Edd}$, the disk will be more
prominent.

Now we can calculate a synthetic log $N$ - log $L$ distribution for
our model.  The emission from the disk and the jet is described by
Eq. (\ref{emitlaw}). To reach a given luminosity $L$, only the
difference $L-L_{\rm{disk}}$ has to be reached by the jet due to
boosting.  The estimated number of sources with a luminosity greater
than $L$ is given by:
\begin{eqnarray}
N(L) & = & \sum_{i=N,B} 
\mathcal{N}_i \int dM \int d\dot{M} \mathcal{V}_i(M) \mathcal{W}_i(\dot{M})
 \cdot \nonumber \\
& & \ \ \ \ \ P(L-L_{\rm{disk}}(\dot{M}),L_{\rm{jet}}(\dot{M})) \label{eqn1}
\end{eqnarray}
where the sum goes over the two populations.  

\subsection{The Disk-Only Model}
Of course, with the model discussed above we can also investigate the
alternative scenario, that the soft X-ray emission originates only
from the accretion disk, by setting $\eta = 0$. In this case we have
to leave the upper end of the black hole mass distribution and its
power law index as a free parameter to obtain the high luminosities
observed, and we have $d N/dM=\mathcal{V} (M)\propto M^\zeta$.

With the isotropic disk emission the estimated number of sources with
a luminosity greater than $L$ is given by:
\begin{eqnarray}
N(L) & = & \sum_{i=N,B} 
\mathcal{N}_i \int dM \int d\dot{M} \mathcal{V}_i(M) \mathcal{W}_i(\dot{M})
 \cdot \nonumber \\
& & \ \ \ \ \ \Theta(L-L_{\rm{disk}}(\dot{M})) \label{eqnHMBH}
\end{eqnarray}
where $\Theta$ is the usual step-function. 

\section{Data}
To put meaningful constraints on a beaming model it is essential to
compare the low-luminosity (un-beamed) parent population with the
high-luminosity (beamed) population. However, the published X-ray
population of a single galaxy has only marginal statistics in the
high-luminosity regime.  To get a more general X-ray population in the
low-luminosity regime up to $\approx 5 \cdot 10^{38}
\mbox{erg}/\mbox{s}$, we combine data from the galaxies M101, M31 and
M82 (\citeNP{PenceSnowdenMukai2001}; \citeNP{StefanoKongCarcia2001};
\citeNP{GriffithsPtakFeigelson2000}).  These are three close ($D<10$
Mpc) galaxies with good published Chandra data. To get better statistics
at higher luminosities, we used the luminosity function compiled by
\citeN{RobertsWarwick2000} from ROSAT data of 49 spiral
galaxies from the XHFS-sample.  The host galaxy types are spiral with
the exception of the irregular M82.  However this does not seem
to make a significant difference (see Fig.~1 \& 2).  

To avoid incompleteness near the detection limit, for each data set we
only use X-ray sources with a luminosity of ten times the respective
detection threshold. 
Chandra has a different bandpass (0.3-10 keV) than ROSAT (0.1-2.2 keV),
so we extrapolated the ROSAT-luminosities to the 2-8 keV band. 
If the photon index is not fitted directly we used a
common powerlaw with $\Gamma_{\rm ph}=1.7$.  In some cases different
values for $N_H$ were used which we did not correct. According to Di
Stefano et al.  \cite{StefanoKongCarcia2001}, 
variations of $N_H = 0.6 - 1.5 \cdot
10^{21} /\mbox{cm}^2$ and $\Gamma_{\rm ph} = 1.2 - 2$ give differences
in luminosity of about $20\%$
which are not really significant in the
log(N)-log(S) plots shown here and should be statistically
distributed (for ROSAT data the differences are higher).

As a reference galaxy we take M101, to which we scale the populations
of the other galaxies (i.e., the total number of sources in the overlapping 
luminosity bins). 
Combining these data sets assumes that the overall shape of the
luminosity distribution is roughly universal. Clearly, the overall
number of XRBs in each galaxy can depend strongly on the age of recent
star formation, but the average slope of the luminosity function
should be less sensitive to this.  Since M82 is irregular and has a
much higher star forming rate than M101 or M31, we also show the data
excluding M82, which is not significantly different.

To calculate errors we assume a standard deviation from the 'general
population' of $\sqrt{N}$ where N is the number of detected sources,
and use normal error propagation.  Because we are showing a cumulative
distribution, the errors for each point are not independent.

\section{Results}
To compare our simple model with the data we evaluate the integrals in
Eqs. (\ref{eqn1} \& \ref{eqnHMBH}) numerically. The absolute
normalization and the parameter $\xi$, which are entirely free, have
been fit to the data at $L_{\rm x}\le10^{37} \mbox{erg} \
\mbox{s}^{-1}$. We obtain a best-fit value of the accretion rate index
$\xi = 1.4$ (note that the luminosity scales as $\dot M^2$ in this
regime). As we only model neutron stars and black holes, the model fits 
could be affected at lower luminosities by other
source populations like accreting white dwarfs and supernova remnants.

\begin{figure}
\resizebox{\hsize}{!}{\includegraphics{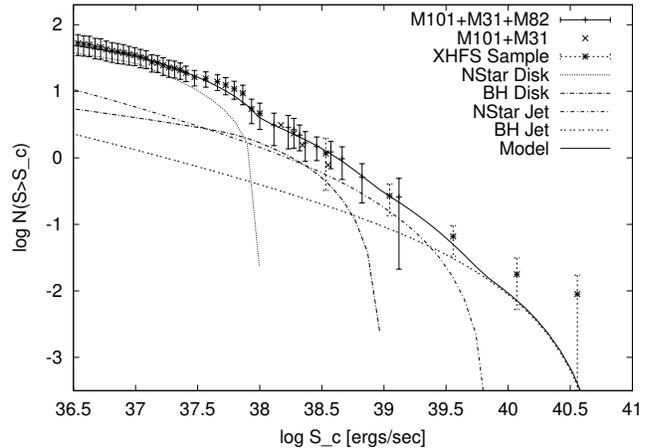}}
\caption{Comparison of our model of the luminosity function with the data.
The parameters are $\gamma_{\rm j} = 5$ , $\eta = 0.3$, $\dot{M}_C = 0.1$, $\xi =1.4$.
Also shown are the individual contributions of the disk and the jet
for neutron stars and black holes. }
\label{gamma5}
\end{figure}

Fig. ~\ref{gamma5} shows the result for our best-fit jet/disk model
which requires $\gamma_{\rm j} = 5$ and $\eta = 0.3$, for the combined
data set discussed above. The Eddington limit for black holes (limited
to $M<15M_\odot$) and neutron stars shows up as breaks at the
respective luminosities, and jets with $\gamma_{\rm j} = 5$ are able
to produce emission up to $10^{40} \mbox{erg} \ \ \mbox{s}^{-1}$ in
significant numbers.
The model is most sensitive to $\gamma_{\rm j}$ and $\eta$.
Because the high luminosity domain depends linearly on $\eta$ while
its dependence on $\gamma_{\rm j}$ goes as $\gamma_{\rm j}^{2.7}$, a
slight decrease of $\gamma_{\rm j}$ can be compensated by an increase
of $\eta$ and vice versa. For $\gamma_{\rm j}=5.8 $ or $\gamma_{\rm
j}=7.5$ we can find $\eta=0.2$ or $\eta=0.1$, but the fit gets
progressively worse at higher Lorentz factors. Demanding $\eta\la0.3$
for the radiative efficiency of the jet sets a lower limit for
$\gamma_{\rm j}\ga5$.

\begin{figure}
\resizebox{\hsize}{!}{\includegraphics{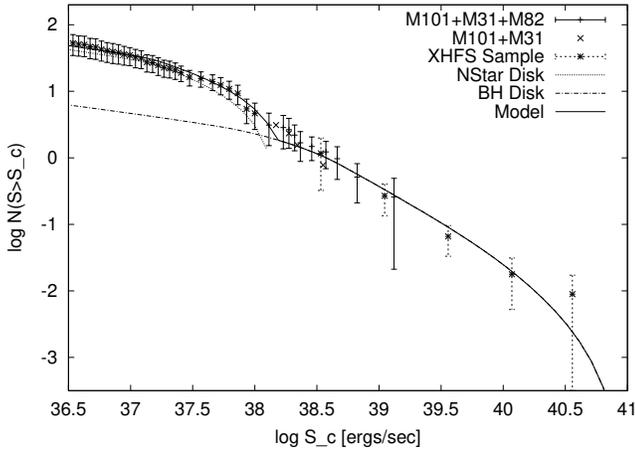}}
\caption{Model with intermediate-mass black holes up to $1000 M_\odot$ 
and a mass distribution with power law index $\zeta=2$.}
\label{intermed}
\end{figure}

For the disk-only model, the sensitive parameters are the power law
indices of the accretion rate and the mass distribution of the black
holes. To fit the data, a mass distribution index of $\zeta\simeq2$ is
needed. The index of the accretion rate is the same as before (1.4),
because the lower luminosities in the jet/disk model are also
dominated by the disk. To explain the most luminous sources, the upper
end of the black hole mass distribution must be extended at least up
to 1000 $M_\odot$. The results of the fit are shown in
Fig.~\ref{intermed}.  For the lower luminosities the neutron stars
dominate the luminosity function, while the black holes dominate at
higher luminosities.

\section{Summary and Discussion}
Using two very simple models for the evolution of XRBs, we calculate
the luminosity distribution of X-ray point sources in nearby
galaxies. We consider a jet/disk model based on \citeN{FalckeBiermann1999} 
and \citeN{MarkoffFalckeFender2001},
which can give rise to relativistically beamed emission from
microblazars. Alternatively we also consider a purely isotropically
radiating disk model.

Both models can in principle reproduce a combined luminosity function
compiled from X-ray point source catalogs of three close galaxies and
the XHFS spiral galaxy sample.  However, as expected, the isotropic
disk model requires a mass distribution of black holes extending out
to $1000 M_\odot$ to explain the ULXs. On the other hand, a
relativistic jet/disk model can fit the data with stellar mass black
holes, if X-ray emitting jets with Lorentz factors $\gamma_{\rm
j}\simeq5$ are present in XRBs. In addition, a fraction of
$\eta=10-30\%$ of the total soft X-ray emission has to come from the
jet rather than the accretion disk for an un-beamed XRB in the high
state. This requires rather powerful jets but is not completely
unreasonable. If only a fraction of the XRBs have relativistic jets,
a slightly higher Lorentz factor or jet efficiency is needed.
Boosting a $10$ mJy Galactic XRB by a factor $\sim 10^2$ (for $\gamma \sim 5$) 
and placing it
at $D \sim 3 {\rm Mpc}$ would yield only a faint $10$ nJy source
and make radio detections difficult. 

With the current statistics it is not possible to distinguish between
the two different models, but it seems that microblazars provide at
least a sensible alternative to the often discussed intermediate mass
black hole scenario. Monitoring the spectral variability of the most
luminous sources and further developing the XRB jet model should
eventually help to disentangle the two scenarios.

\vspace{3mm}

\noindent {\it Acknowledgments}
The authors would like to thank R. Fender and an anonymous referee for
useful suggestions.
%\bibliography{aamnemonic,Dh271_refs}
%\bibliographystyle{aa}

\end{document}